# Extending Ambient Pressure X-ray Photoelectron Spectroscopy to Plasma Studies:

## A novel and flexible plasma gun approach

Yang Gu[1,2,#], Zhehao Qiu[1,#], Shui Lin[1,2], Yong Han[2], Hui Zhang[3], Zhi Liu[1,2,*], Jun Cai[1,2,*]

1 School of Physical Science and Technology, ShanghaiTech University, Shanghai 201210, China;

2 Center for Transformative Science, ShanghaiTech University, Shanghai 201210, China;

3 Shanghai Synchrotron Radiation Facility Shanghai Advanced Research Institute Chinese Academy of Sciences, Shanghai 20124, China

[#] These authors contributed equally.

*Corresponding author:

Zhi Liu: liuzhi@shanghaitech.edu.cn

Jun Cai: caijun@shanghaitech.edu.cn

**Abstract**

The characterization of the electronic structure and chemical states of gases, solids, and liquids can be effectively performed using ambient pressure X-ray photoelectron spectroscopy (AP-XPS). However, the acquisition of electronic and chemical information under plasma conditions poses significant challenges. In this study, we have developed an advanced experimental system capable of garnering electronic information amidst plasma environments, alongside providing detailed surface chemical states of samples subjected to plasma conditions. By designing a customized plasma generation apparatus, we successfully integrated it with a traditional AP-XPS system. This novel plasma-AP-XPS system confined plasma proximal to the sample area, with adjustable intensity parameters controlled by either modifying the distance between the plasma source and the sample surface or adjusting the voltage applied. This configuration permitted the direct detection of electrons in the plasma via the XPS electron detector. To substantiate the efficacy and versatility of this setup, it was applied to two distinct studies: the plasma etching of graphene and plasma oxidation of platinum (Pt). The investigations confirmed that argon (Ar) plasma facilitates the etching of graphene, a phenomenon clearly evidenced by the XPS spectra. Similarly, the exposure of the Pt surface to oxygen plasma was found to induce effective oxidation. This developed system significantly extends the utility of AP-XPS, enhancing its application for in-depth studies of plasma-enhanced reactions under operando conditions, thereby holding promise for the advancement in material science and chemical engineering fields.

**Introduction**

X-ray photoelectron spectroscopy (XPS) is a highly versatile and powerful technique in the realm of surface science, known for its capacity to deliver comprehensive information regarding the surface composition, chemical states, and electronic environments of a wide range of materials.[1-3] Through the measurement of the kinetic energy of photoelectrons ejected from the material surface under X-ray irradiation, XPS provides a non-destructive method for probing the atomic and electronic structures within the uppermost few nanometers of a specimen.[4] Traditionally, XPS experiments are carried out under ultra-high vacuum (UHV) conditions. This environment is crucial to ensure the accurate detection of photoelectrons, eliminating potential interferences from gas-phase or liquid-phase species.[5, 6] These ex-situ XPS measurements (i.e., those performed under UHV conditions) have significantly advanced our understanding of various material properties. However, a major limitation of traditional XPS arises from its constraint to UHV environments. This restriction impairs the technique's applicability to real-world processes, as numerous critical surface reactions occur under non-vacuum conditions. Specifically, those reactions take place in the presence of plasma conditions, the fourth state of matter.

To explore the surface states under reaction conditions, ambient pressure X-ray photoelectron spectroscopy (AP-XPS, also referred to as in-situ XPS) techniques have been developed. These techniques enable the characterization of materials under more realistic conditions, including elevated pressures and reactive environments. Over decades of dedicated research and development, the working pressure limits of AP-XPS has been significantly expanded to the realm of hundreds of millibars. This expansion is attributed to the design of differential pumping systems and the optimization of front cone components.[4, 7-10] The rapid advancements in AP-XPS experimental setups have significantly enhanced our detection capabilities, providing profound insights into material properties under gaseous or liquid conditions.[5, 11-15]

Among these advancements, plasma AP-XPS represents a significant leap forward, as it enables the direct observation of surface chemistry under plasma conditions.[16] Plasmas, inherently composed of a myriad of reactive entities including ions, electrons, and radicals, are pivotal in a broad spectrum of industrial processes, ranging from plasma catalysis reaction[17, 18], semiconductor manufacturing[19] to thin-film deposition[20], and surface etching[21]. The ability to monitor surface reactions in real time during plasma exposure is therefore critical for the enhancement of these technological processes and the comprehensive understanding of their foundational mechanisms. Integrating the AP-XPS methodology with a dynamic plasma environment, this technique offers a singular research platform to probe the complex interactions between plasma-induced species and material surfaces.[22] This integration, however, presents several technical challenges. The intensity of the plasma must be calibrated to preclude potential damage to the $Si_3N_4$ window—a critical component that separates the X-ray source and the main chamber, with a thickness of 100 nanometers. Additionally, the presence of plasma potentially introduces extraneous noise and the accumulation of charges on the sample surface, necessitating an optimization of the experimental configuration. Beyond these considerations, achieving a comprehension of how plasma parameters—including power, pressure, and gas composition—impact XPS measurements is essential for the accurate interpretation of experimental results.

In this work, we report a comprehensive investigation into the development and application of in-situ plasma XPS. We begin by discussing the design and optimization of the experimental setup, focusing on strategies to mitigate the challenges associated with plasma environments. The electrons in the plasma could be characterized by the XPS electron detector. Then we explore the application of in-situ plasma XPS to a range of material systems, highlighting its ability to reveal the real-time evolution of surface chemistry during plasma processes. Our investigation focuses on emblematic case studies—plasma oxidation, reduction, and etching processes—to concretize the potential of this methodology. These case studies serve to illuminate how in-situ plasma XPS can provide crucial insights into the mechanisms driving plasma-surface interactions, which are marked by the simultaneous occurrence of multiple, conflicting pathways. Moreover, we discuss the implications of our findings for advancing plasma-based technologies, particularly in the surface etching and oxidation, where precise control over surface properties is essential. The ability to monitor and understand plasma-induced surface modifications at the atomic level opens up new possibilities for tailoring material properties to meet specific application requirements.

**Design of the plasma setup**

Figure 1 shows the overview of the plasma generation setup. The plasma was generated by applying a direct current (DC) electric field between two electrodes within a gas-filled conditions. The anode and cathode electrodes both were made from tungsten, which could withstand the harsh conditions of plasma discharge, including high temperatures, ion bombardment, and potential chemical reaction with the plasma species. The electrodes were capsuled by ceramic to insulate them. The gases were introduced into the chamber through the cylindrical region between the anode and cathode electrodes. Upon reaching a critical voltage threshold, the gas between the electrodes could be ionized, forming the plasma-a partially ionized gas composed of ions, electrons, and neutral atoms or molecules. Additionally, the whole plasma generator were connected with a linear driver and a port aligner to scale freely.

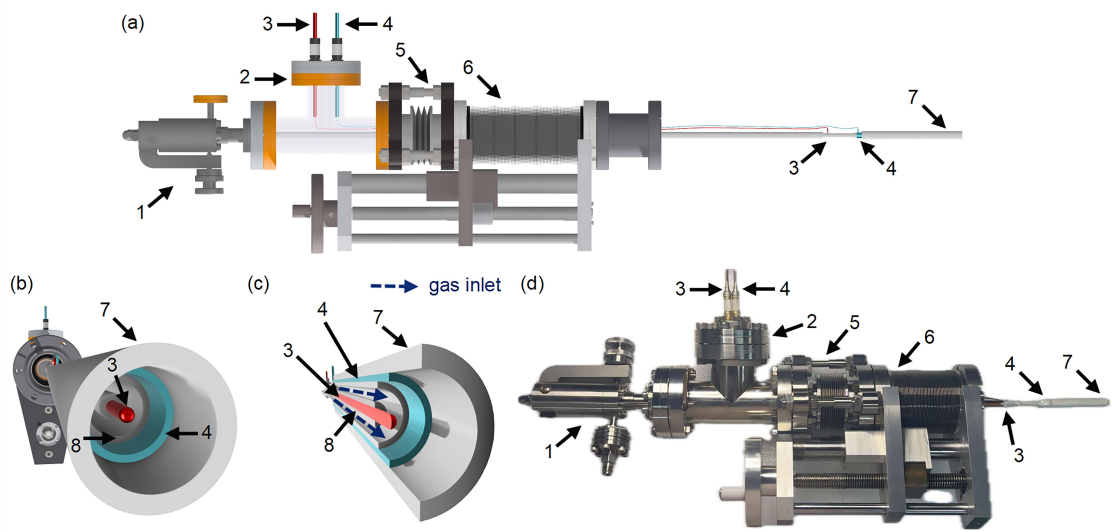

Figure 1 Overview of the plasma generation setup. (a) The scheme of the DC plasma gun; (b) Inside of the plasma gun; (c) Cross section view of the plasma gun; (d) The image of the plasma gun. It consists of (1) leak valve, (2) power through, (3) anode, (4) cathode, (5) port aligner, (6) linear driver, (7) plasma nozzle, (8) insulation layer.

Figure 2 shows the picture of the plasma setup in the AP-XPS system. There are several parts in the system, including the main chamber, differential pumps, analyzer, X-ray source, and the preparation chamber. The plasma setup was connected in the main chamber (Figure 2(a)). The sample was weld in the sample holder with an inconel foil. The plasma spray nozzle can move close to the sample's surface. Due to the geometric constraints of the plasma spray nozzle, the minimum distance from the top of the spray nozzle to the sample surface is approximately 5 mm, thereby confining the plasma to a localized region near the material. An angular arrangement has been optimized such that the X-ray $Si_3N_4$ window and the plasma spray nozzle form an approximate 45-degree angle, strategically directing the plasma flux away from the window to prevent potential damage and contamination. Figure 2(d-f) shows the visually distinct plasmas generated with different gases — carbon dioxide ($CO_2$), oxygen ($O_2$), and nitrogen ($N_2$) — each exhibiting unique color characteristics.[23] Such visual evidence represents the diverse plasma conditions that can be achieved and studied within this experimental setup, providing ample opportunities for tailoring plasma gas composition according to the specific requirements of the surface reactions. This tailored approach enhances the adaptability of the system to a wide range of applications, from material modification to catalysis research.

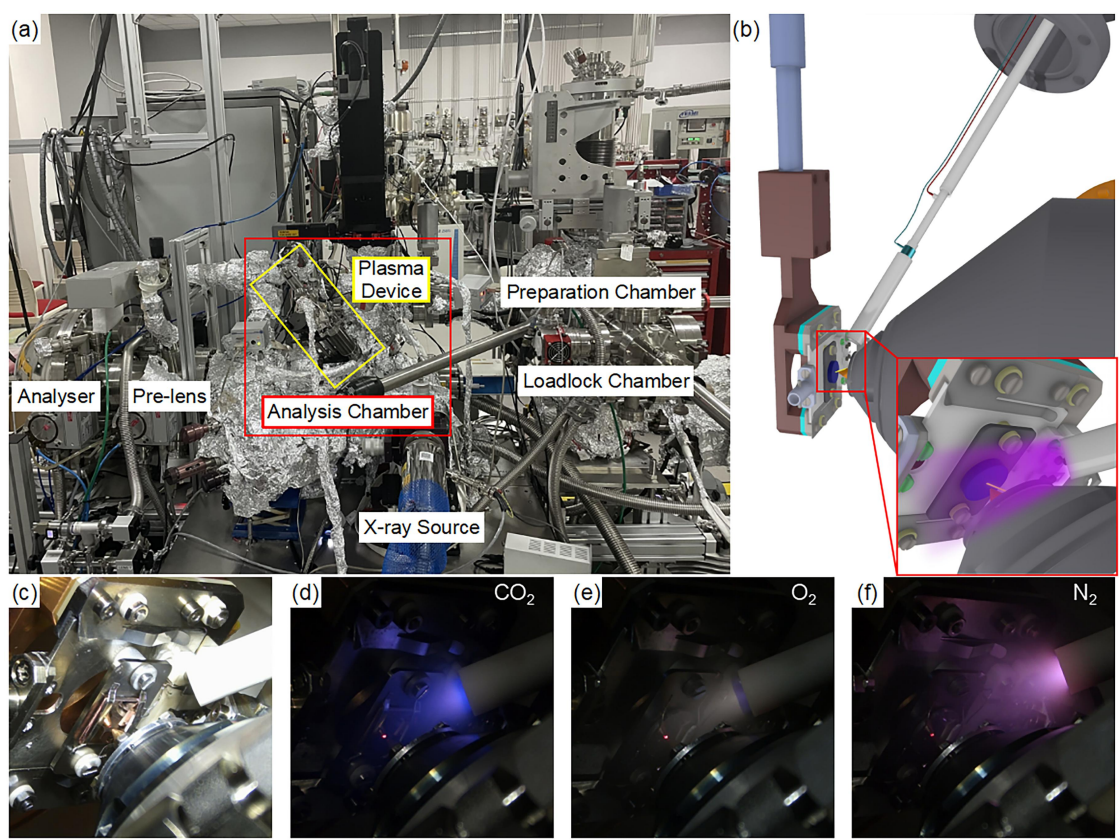

Figure 2 The plasma setup in the main chamber. (a) The lab-based AP-XPS system, the yellow box is the plasma generation part; (b) The scheme of the inside of the main chamber; (c) The picture of sample in the main chamber; (d-f) The picture of the $CO_2$, $O_2$ and $N_2$ plasma in the main chamber,

respectively.

As the plasma consists of ions, electrons, radicals, and other reactive species.[24] We can use the electron analyzer to detect the electrons in the plasma. The corresponding results are shown in Figure 3, which delineates the spectra of the electrons in the $N_2$ plasma. It should be noted that the X-ray source is off and there is no sample in the main chamber during the spectra collection. The $N_2$ plasma was generated by apply 600 V voltage between the two electrodes. The pass energy was set to the lowest 5 eV to collect electrons in the maximum kinetic energy range. Most of the electrons' energy are concentrated around 11.5 eV (Figure 3(a,b)). This observation indicates that the average energy of the electrons in the plasma is about 11.5 eV. And there are three peaks at the higher kinetic energy part. There are three distinct peaks in the higher kinetic energy region. Specifically, the separation between these adjacent peaks approximates 13.5 eV (Figure 3(c)). This separation correlates directly with the process where electrons, initially possessing an energy of 600 eV and devoid of scattering, engage in interactions with gas molecules. Notably, this energy differential is closely aligned with the energy required for the first ionization of $N_2$, which is approximately 13.5 eV.[25] The third peak represents that the electrons with an energy of 600 eV excited the $N_2$ molecule two times. Analogous observations are made for other gases, such as $O_2$ and $CO_2$, where the peak separation represents the energy requisite for the first ionization process of respective gas (Table 1). Consequently, the peaks at the higher kinetic energy part are the characteristic peaks of the different kind of gas plasma.

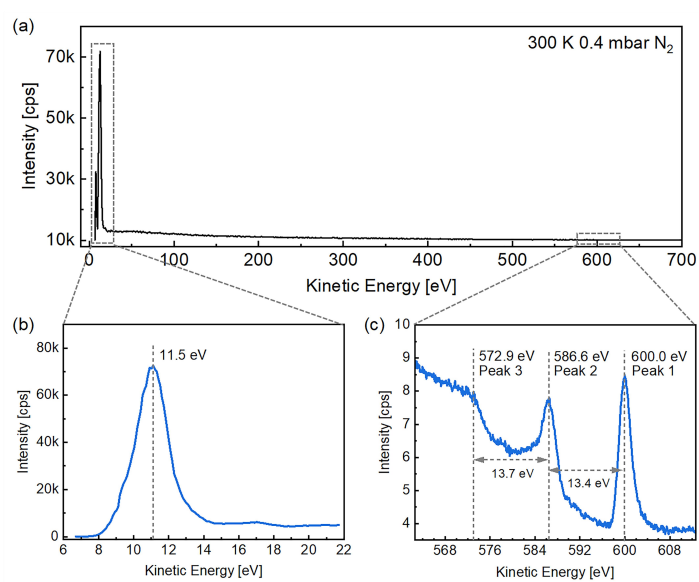

Figure 3 Spectra of $N_2$ plasma. (a) Survey of the electron in the $N_2$ plasma, the spectrum was collected with a pass energy of 5 eV; (b) The fine spectrum in the low electron energy part; (c) The fine spectrum in the high energy part. The plasma was generated by apply 600 V voltage between the two electrodes.

Table 1. XPS peak separation of different gases under plasma condition.

| Gas | $\Delta_{Peak\ 1,2}$ (eV) | $\Delta_{Peak\ 2,3}$ (eV) |
| --- | --- | --- |
| $N_2$ | 13.4 | 13.7 |
| $O_2$ | 8.5 | 8.3 |
| $CO_2$ | 12.7 | 13.1 |

A DC voltage was applied to induce a discharge within the inter-electrode gap, which was filled with gas. The electrons from the negative cathode were accelerated in the electric field established across the gas-filled interstice that separated the cathode from the anode.[23] The discharge type depends on the working gas pressure, the applied voltage and the geometry of the discharge.[26] In this work, we studied the properties of the plasma in the DC glow discharge region. The low energy electrons at around 11 eV were inferred to originate from secondary electronic emissions. It was observed that the kinetic energy distribution of low energy electrons remained invariant as the DC voltage changes, as shown in Figure 4(c). The secondary electron energy depends on the kinetic transfer of the potential energy from the incoming ions or neutral atoms to the electrons in the cathode surface, a mechanism analogous to the auger process.

To study the plasma electron density as a function of distance, we take the spectra of the low energy electrons while changing the nozzle distance, as shown in Figure 4(a). It can be seen that the characteristic peak kept at the same position at different distance (Figure 4(b)). This constant peak position is accompanied by a decrease in peak intensity, a trend that corresponds with the visual representation of plasma in the right part of Figure 4(b). The integration of the spectra was shown in the inset of the Figure 4(b), indicating that the electrons were scattered by $N_2$ in the chamber.

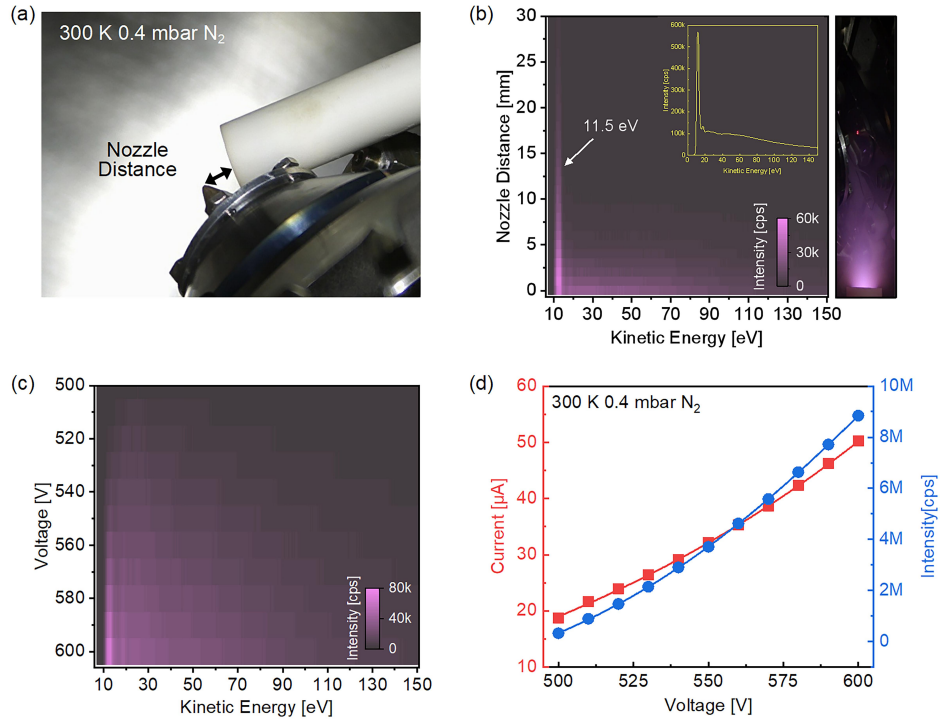

Figure 4 The spectra collected under different conditions. (a) The distance from the plasma nozzle to the XPS cone; (b) The left part is the low energy electron spectra as a function of the nozzle distance, the wright part is the image of the plasma in the chamber; (c) The spectra collected under different DC voltage; (d) The variation of the plasma current and intensity as the DC voltage increases.

The influence of direct current (DC) voltage on plasma intensity was evaluated, with results shown in Figure 4(c) representing that the intensity of the electron kinetic energy decreases as the DC voltage decreases. The plasma current between the anode and cathode electrodes was measured by a Keithley meter. As shown in Figure 4(d), the measurements were in constant with the observed variations in plasma intensity, further demonstrating the direct impact of DC voltage modulation on plasma characteristics.

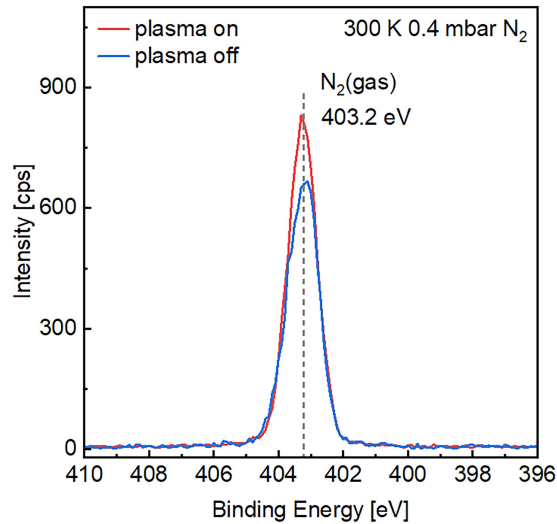

Figure 5 The spectra of $N_2$ collected under the plasma on and off, respectively.

In the configuration of the plasma generation apparatus, an electric field is established between the anode and cathode, which may potentially influence the acquisition of the XPS spectra. To study the presence and potential impact of the electrical field near the sample, the XPS spectra of the $N_2$ gas phase were collected under the condition where the plasma activated ('on') and deactivated ('off'). Figure 5 shows that the peak position of the $N_2$ gas phase under the condition of plasma on or off keeps at the same position. This suggests the absence of an electric field near the sample during plasma activation, providing critical insights into the characteristics of the plasma generation setup and its influence on XPS spectral collection.

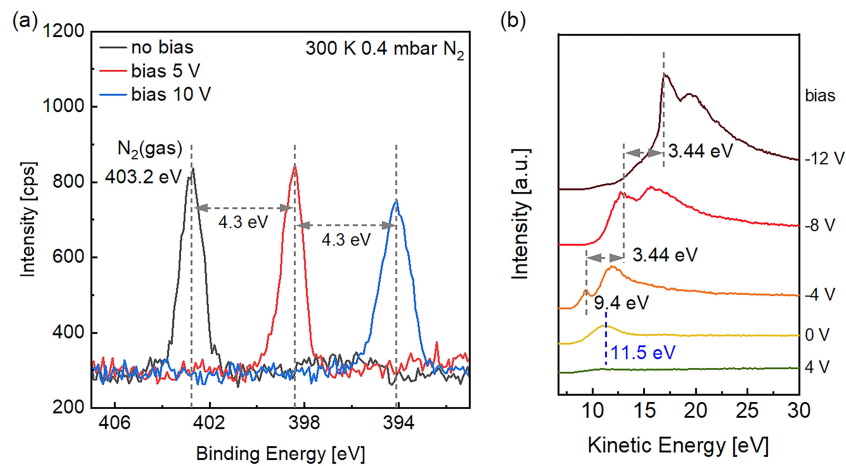

Figure 6 (a) The spectra of $N_2$ gas phase under different bias; (b) The spectra at the lower kinetic energy part collected at different bias.

To detect the lower energy electrons, we applied variable bias on the sample at the test position. As shown in Figure 6(a), the $N_2$ gas phase shifted to lower binding energy as we applied positive bias on the sample. An electric field between the sample and the analyzer cone could trigger off the acceleration of electrons with kinetic energy below 10 eV upon the application of a negative bias, thereby enabling their passing into the analyzer. When we applied positive bias, the lower kinetic electrons were attracted and the electron intensity decreased significantly, as shown in Figure 6(b). We can see that the electrons with kinetic energy bellow 11.5 eV could be detected when a negative bias was applied on the sample. The characteristic peak is around 5.4 eV after correction for the applied bias. This methodology emerges as a viable technique for the detection of low-energy electrons under plasma conditions. And the origin and principle of the electrons needs further exploration.

**Experimental cases**

(a) Plasma etching

Plasma-assisted etching has becomed an indispensable technique in contemporary semiconductor manufacturing. To study the plasma etching process, we used the graphene covered Rh(111) as a model system. The sample was mounted on the sample holder and the plasma nozzle was near the sample to ensure effective interaction during the etching process. The graphene was grown on the Rh(111) surface by chemical vapor deposition, using a mixture gas of $C_2H_4$ and $H_2$ at 1000 K.[27] And the as-prepared graphene/Rh(111) was exposed to the 0.4 mbar Ar environment. Then we turn on plasma and collect the C 1s spectra on the sample surface at room temperature. Figure 7(a) illustrates the time-dependent decrease of the C 1s peak intensity of graphene on Rh(111) throughout the etching process, highlighting the complete disappearance of graphene after a period of 20 minutes. And the etching rate was shown in Figure 7(b). By positioning the plasma nozzle at various distances to the sample surface, we can see the etching rate decreases as the distance increases.

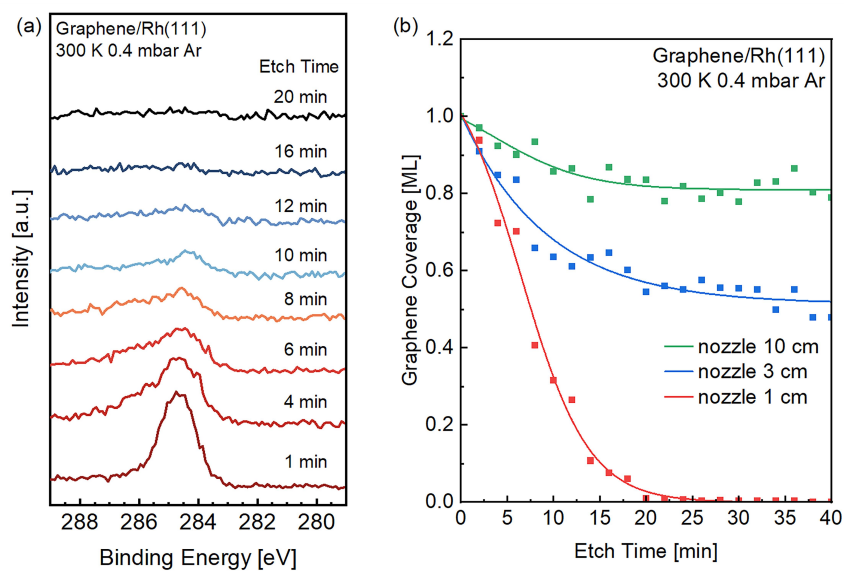

Figure 7 Plasma etching of the graphene on Rh(111). (a) The evolution of C 1s spectra of graphene on Rh(111) collected at room temperature under the Ar plasma environment; (b) The variation of the C 1s intensity of graphene during the etching process at different distance.

(2) Surface oxidation under plasma conditions

Plasma, a highly energized state of matter, is composed of ions, electrons, radicals, and a multitude of reactive species. The interaction of these constituents with material surfaces could induce complex chemical reactions, significantly altering the surface chemical states and structures.[28, 29] To study the plasma enhanced chemical reaction on the metal surface, we use the Pt(111) oxidation by $O_2$ as an example. The XPS spectra on the Pt(111) surface were shown in the following Figure 8.

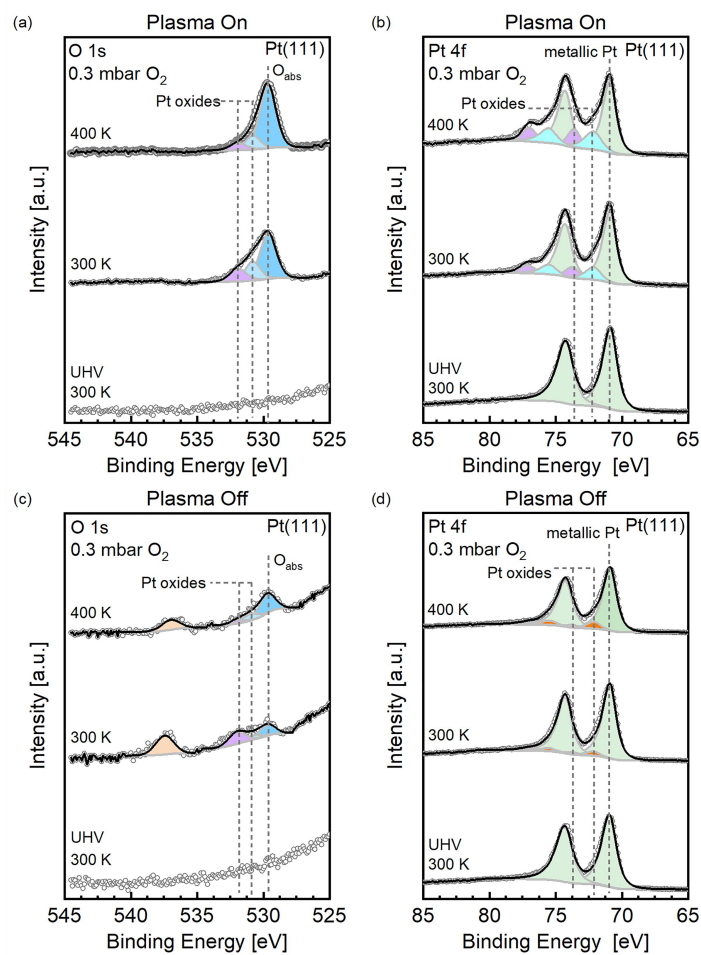

Figure 8 XPS spectra on the Pt(111) surface. (a) O 1s and (b) Pt 4f spectra collected under the oxygen plasma conditions (0.4 mbar $O_2$) from 300 K to 400 K; (c) O 1s and (d) Pt 4f spectra under 0.4 mbar $O_2$ on the Pt(111) surface with plasma off.

The Pt(111) sample was cleaned by cycles of sputtering and annealing, as verified by the XPS spectra in Figure 8. After the cleaning procedure, 0.3 mbar $O_2$ was introduced into the chamber at room temperature. And then we turn on the $O_2$ plasma. Figure 8 shows the O 1s and Pt 4f spectra on the Pt(111) surface. The surface was covered with adsorbed oxygen (529.7 eV), Pt oxide (surface oxide: 530.9 eV, $PtO_2$: 531.6 eV) at 300 K.[30, 31] In the Pt 4f spectra, there is also the corresponding peaks of the Pt-O species (adsorbed oxygen and surface oxide/$PtO_2$).[31-33] As the Pt(111)'s temperature increases to 400 K, the surface was further oxidized as shown in Figure 8(a,b). The gas peak under plasma conditions was very weak at the same pressure. The Pt surface was oxidized.

In contrast, we collected the O 1s and Pt 4f spectra on Pt(111) surface in oxygen conditions with plasma off. The corresponding spectra were shown in Figure 8(c,d). The results demonstrated a notable difference between the spectra collected under $O_2$ conditions and those during $O_2$ plasma exposure. The enhancement of oxidation processes attributable to the plasma influence could thus be confirmed.

**Conclusion**

In summary, we show an advanced experimental system that combines the plasma generator with the AP-XPS. The plasma can be confined near the sample region in the main chamber, allowing for the direct characterization of electrons in the plasma utilizing the electron energy analyzer of the XPS system. The innovative feature of this setup is the ability to collect XPS spectra under plasma conditions, known as plasma AP-XPS. This methodology holds potential for a diversity of plasma-involved processes, including plasma catalysis, semiconductor manufacturing, thin-film deposition, and precise surface etching. It represents a powerful tool for the real-time investigation of surface reactions under the plasma exposure, offering new insights into the dynamic processes that govern material behaviors in industrially relevant environments. The ongoing refinement and application of in-situ plasma XPS are poised to significantly enrich our comprehension of the plasma-assisted material processing, potentially catalyzing the advent of cutting-edge technological innovations.

**Acknowledge**

This work was mainly supported by the National Key R&D Program of China No. 2022YFA1503802. This work was supported by Shanghai Rising-Star Program No. 21QA1406200 and the National Natural Science Foundation of China 22072093, 21991152 and 22002090. The authors thank the support of BL02B01 of the Shanghai Synchrotron Radiation Facility and SPECS AP-XPS instrument supported by the National Natural Science Foundation of China, No.11227902.

**Author Contributions**

Zhi Liu conceived the research and supervised the project. Jun Cai and Yang Gu designed this plasma gun. The experiments were performed by Yang Gu, Zhehao Qiu and Jun Cai. Shui Lin helped to perform the experiment. Yong Han and Hui Zhang helped to analyze the experiment results. Yang Gu and Zhehao Qiu wrote the original draft. Important contributions to the interpretation of the results, conception and revise of the manuscript were made by Jun Cai and Zhi Liu. All authors participated in the scientific discussion.

**Notes**

The authors declare no competing financial interest.

**Reference**

(1) Fadley, Cs. X-ray photoelectron spectroscopy: Progress and perspectives. *Journal of Electron Spectroscopy and Related Phenomena* **2010**, *178*, 2-32.

(2) Salmeron, Miquel; Schlögl, Robert. Ambient pressure photoelectron spectroscopy: A new tool for surface science and nanotechnology. *Surface Science Reports* **2008**, *63* (4), 169-199.

(3) Siegbahn, K; Nordling, C; Johansson, G; Hedman, J; Hedén, Pf; Hamrin, K; Gelius, U; Bergmark, T; Werme, Lo; Manne, R. ESCA Applied to Free Molecules North-Holland Publ. *Co.(Amsterdam)* **1969**.

(4) Cai, Jun; Dong, Qiao; Han, Yong; Mao, Bao-Hua; Zhang, Hui; Karlsson, Patrik G.; Åhlund, John; Tai, Ren-Zhong; Yu, Yi; Liu, Zhi. An APXPS endstation for gas–solid and liquid–solid interface studies at SSRF. *Nuclear Science and Techniques* **2019**, *30* (5), 81.

(5) Starr, De; Liu, Z; Hävecker, Michael; Knop-Gericke, Axel; Bluhm, H. Investigation of solid/vapor interfaces using ambient pressure X-ray photoelectron spectroscopy. *Chem. Soc. Rev.* **2013**, *42* (13), 5833-5857.

(6) Zhang, Hui; Li, Xiaobao; Wang, Wei; Mao, Baohua; Han, Yong; Yu, Yi; Liu, Zhi. Ambient pressure mapping of resonant Auger spectroscopy at BL02B01 at the Shanghai Synchrotron Radiation Facility. *Review of Scientific Instruments* **2020**, *91* (12), 123108.

(7) Ogletree, D. Frank; Bluhm, Hendrik; Lebedev, Gennadi; Fadley, Charles S.; Hussain, Zahid; Salmeron, Miquel. A differentially pumped electrostatic lens system for photoemission studies in the millibar range. *Review of Scientific Instruments* **2002**, *73* (11), 3872-3877.

(8) Frank Ogletree, D.; Bluhm, Hendrik; Hebenstreit, Eleonore D.; Salmeron, Miquel. Photoelectron spectroscopy under ambient pressure and temperature conditions. *Nuclear Instruments and Methods in Physics Research Section A: Accelerators, Spectrometers, Detectors and Associated Equipment* **2009**, *601* (1), 151-160.

(9) Grass, Michael E; Karlsson, Patrik G; Aksoy, Funda; Lundqvist, Måns; Wannberg, Björn; Mun, Bongjin S; Hussain, Zahid; Liu, Zhi. New ambient pressure photoemission endstation at Advanced Light Source beamline 9.3.2. *Review of Scientific Instruments* **2010**, *81* (5), 053106.

(10) Amann, Peter; Degerman, David; Lee, Ming-Tao; Alexander, John D.; Shipilin, Mikhail; Wang, Hsin-Yi; Cavalca, Filippo; Weston, Matthew; Gladh, Jörgen; Blom, Mikael; Björkhage, Mikael; Löfgren, Patrik; Schlueter, Christoph; Loemker, Patrick; Ederer, Katrin; Drube, Wolfgang; Noei, Heshmat; Zehetner, Johann; Wentzel, Henrik; Åhlund, John; Nilsson, Anders. A high-pressure x-ray photoelectron spectroscopy instrument for studies of industrially relevant catalytic reactions at pressures of several bars. *Review of Scientific Instruments* **2019**, *90* (10), 103102.

(11) Han, Yong; Zhang, Hui; Yu, Yi; Liu, Zhi. In Situ Characterization of Catalysis and Electrocatalysis Using APXPS. *ACS Catal.* **2021**, 1464-1484.

(12) Eriksson, Susanna K; Hahlin, Maria; Kahk, Juhan Matthias; Villar-Garcia, Ignacio J; Webb, Matthew J; Grennberg, Helena; Yakimova, Rositza; Rensmo, Håkan; Edström, Kristina; Hagfeldt, Anders. A versatile photoelectron spectrometer for pressures up to 30 mbar. *Review of Scientific Instruments* **2014**, *85* (7), 075119.

(13) Axnanda, Stephanus; Crumlin, Ethan J; Mao, Baohua; Rani, Sana; Chang, Rui; Karlsson, Patrik G; Edwards, Mårten Om; Lundqvist, Måns; Moberg, Robert; Ross, Phil. Using "Tender" X-ray ambient pressure X-Ray photoelectron spectroscopy as a direct probe of solid-liquid interface. *Scientific reports* **2015**, *5*.

(14) Cai, Jun; Ling, Yunjian; Zhang, Hui; Yang, Bo; Yang, Fan; Liu, Zhi. Formation of Different Rh–O Species on Rh (110) and Their Reaction with CO. *ACS Catal.* **2022**, *13* (1), 11-18.

(15) Zang, Yi-Jing; Shi, Shu-Cheng; Han, Yong; Zhang, Hui; Wang, Wei-Jia; Liu, Peng; Ye, Mao; Liu, Zhi. Combination of a reaction cell and an ultra-high vacuum system for the in situ preparation and characterization of a model catalyst. *Nuclear Science and Techniques* **2023**, *34* (5), 65.

(16) Diulus, J. Trey; Naclerio, Andrew E.; Boscoboinik, Jorge Anibal; Head, Ashley R.; Strelcov, Evgheni; Kidambi, Piran R.; Kolmakov, Andrei. Operando XPS for Plasma Process Monitoring: A Case Study on the Hydrogenation of Copper Oxide Confined under h-BN. *J. Phys. Chem. C* **2024**, *128* (18), 7591-7600.

(17) Winter, Lea R.; Chen, Jingguang G. $N_2$ Fixation by Plasma-Activated Processes. *Joule* **2021**, *5* (2), 300-315.

(18) Neyts, Erik C.; Ostrikov, Kostya; Sunkara, Mahendra K.; Bogaerts, Annemie. Plasma Catalysis: Synergistic Effects at the Nanoscale. *Chem. Rev.* **2015**, *115* (24), 13408-13446.

(19) Kanarik, Keren J. Inside the mysterious world of plasma: A process engineer's perspective. *Journal of Vacuum Science & Technology A* **2020**, *38* (3).

(20) Vallée, Christophe; Bonvalot, Marceline; Belahcen, Samia; Yeghoyan, Taguhi; Jaffal, Moustapha; Vallat, Rémi; Chaker, Ahmad; Lefèvre, Gautier; David, Sylvain; Bsiesy, Ahmad; Possémé, Nicolas; Gassilloud, Rémy; Granier, Agnès. Plasma deposition—Impact of ions in plasma enhanced chemical vapor deposition, plasma enhanced atomic layer deposition, and applications to area selective deposition. *Journal of Vacuum Science & Technology A* **2020**, *38* (3).

(21) Donnelly, Vincent M.; Kornblit, Avinoam. Plasma etching: Yesterday, today, and tomorrow. *Journal of Vacuum Science & Technology A* **2013**, *31* (5).

(22) Gerenser, L. J. XPS studies of in situ plasma-modified polymer surfaces. *Journal of Adhesion Science and Technology* **1993**, *7* (10), 1019-1040.

(23) Gudmundsson, Jon Tomas; Hecimovic, Ante. Foundations of DC plasma sources. *Plasma Sources Science and Technology* **2017**, *26* (12), 123001.

(24) Andreev, S. N.; Bernatskiy, A. V.; Ochkin, V. N. The Langmuir probe measurements in a low-pressure discharge supported by hollow cathode using the combined periodic and noise sweep signals. *Vacuum* **2020**, *180*, 109616.

(25) Honda, Masato; Kanai, Kaname; Komatsu, Kenichi; Ouchi, Yukio; Ishii, Hisao; Seki, Kazuhiko. Atmospheric effect of air, $N_2$, $O_2$, and water vapor on the ionization energy of titanyl phthalocyanine thin film studied by photoemission yield spectroscopy. *Journal of Applied Physics* **2007**,

*102* (10).

(26) Benedikt, Jan; Kersten, Holger; Piel, Alexander. Foundations of measurement of electrons, ions and species fluxes toward surfaces in low-temperature plasmas. *Plasma Sources Science and Technology* **2021**, *30* (3), 033001.

(27) Gotterbarm, Karin; Zhao, Wei; Höfert, Oliver; Gleichweit, Christoph; Papp, Christian; Steinrück, Hans-Peter. Growth and oxidation of graphene on Rh (111). *Physical Chemistry Chemical Physics* **2013**, *15* (45), 19625-19631.

(28) Tu, X.; Whitehead, J. C. Plasma-catalytic dry reforming of methane in an atmospheric dielectric barrier discharge: Understanding the synergistic effect at low temperature. *Applied Catalysis B: Environmental* **2012**, *125*, 439-448.

(29) Feng, Jiayu; Sun, Xin; Li, Zhao; Hao, Xingguang; Fan, Maohong; Ning, Ping; Li, Kai. Plasma-Assisted Reforming of Methane. *Advanced Science* **2022**, *9* (34), 2203221.

(30) Saliba, N. A.; Tsai, Y. L.; Panja, C.; Koel, B. E. Oxidation of Pt(111) by ozone (O3) under UHV conditions. *Surf. Sci.* **1999**, *419* (2), 79-88.

(31) Cai, Jun; Wei, Liyang; Liu, Jian; Xue, Chaowu; Chen, Zhaoxi; Hu, Yuxiong; Zang, Yijing; Wang, Meixiao; Shi, Wujun; Qin, Tian; Zhang, Hui; Chen, Liwei; Liu, Xi; Willinger, Marc-Georg; Hu, Peijun; Liu, Kaihui; Yang, Bo; Liu, Zhongkai; Liu, Zhi; Wang, Zhu-Jun. Two-dimensional crystalline platinum oxide. *Nature Materials* **2024**.

(32) Miller, D. J.; Oberg, H.; Kaya, S.; Casalongue, H. S.; Friebel, D.; Anniyev, T.; Ogasawara, H.; Bluhm, H.; Pettersson, L. G. M.; Nilsson, A. Oxidation of Pt(111) under Near-Ambient Conditions. *Physical Review Letters* **2011**, *107* (19), 5.

(33) Butcher, Derek R; Grass, Michael E; Zeng, Zhenhua; Aksoy, Funda; Bluhm, Hendrik; Li, Wei-Xue; Mun, Bongjin S; Somorjai, Gabor A; Liu, Zhi. In situ oxidation study of Pt (110) and its interaction with CO. *J. Am. Chem. Soc.* **2011**, *133* (50), 20319-20325.